\documentclass[a4paper, superscriptaddress, twocolumn, prl
]{revtex4}

\usepackage[colorlinks=true,citecolor=blue,linkcolor=blue,pdfhighlight =/O]{hyperref}
\usepackage{graphicx}
\usepackage{amsmath}
\usepackage{mathtools}
\usepackage{color}
\usepackage{soul}
\usepackage{natbib}

\hypersetup{urlcolor=blue}

\begin{document}

\title{Single Quantum Level Electron Turnstile}

\author{D. M. T. van Zanten}
\affiliation{Universit\'e Grenoble Alpes, F-38000 Grenoble, France}
\affiliation{CNRS, Institut N\'eel, F-38000 Grenoble, France}

\author{D. M. Basko}
\affiliation{Universit\'e Grenoble Alpes, F-38000 Grenoble, France}
\affiliation{CNRS, Laboratoire de Physique et Mod\'elisation des Milieux Condens\'es, F-38000 Grenoble, France}

\author{I. M. Khaymovich}
\affiliation{Universit\'e Grenoble Alpes, F-38000 Grenoble, France}
\affiliation{CNRS, Laboratoire de Physique et Mod\'elisation des Milieux Condens\'es, F-38000 Grenoble, France}
\affiliation{Institute for Physics of Microstructures, Russian Academy of Sciences, 603950 Nizhny Novgorod, GSP-105, Russia }

\author{J. P. Pekola}
\affiliation{Universit\'e Grenoble Alpes, F-38000 Grenoble, France}
\affiliation{CNRS, Institut N\'eel, F-38000 Grenoble, France}
\affiliation{Low Temperature Laboratory, Department of Applied Physics, Aalto University School of Science, FI-00076 Aalto, Finland}

\author{H. Courtois}
\affiliation{Universit\'e Grenoble Alpes, F-38000 Grenoble, France}
\affiliation{CNRS, Institut N\'eel, F-38000 Grenoble, France}

\author{C. B. Winkelmann}
\email{clemens.winkelmann@neel.cnrs.fr}
\affiliation{Universit\'e Grenoble Alpes, F-38000 Grenoble, France}
\affiliation{CNRS, Institut N\'eel, F-38000 Grenoble, France}

\date{\today}

\begin{abstract}


We report on the realization of a single-electron source, where current is transported through a
single-level quantum dot (Q), tunnel-coupled to two superconducting leads (S). 
When driven with an ac gate voltage, the experiment demonstrates electron turnstile operation.
Compared to the more conventional superconductor - normal metal - superconductor turnstile, our
SQS device presents a number of novel properties, including higher immunity to the unavoidable presence of non-equilibrium quasiparticles in superconducting leads. 
In addition, we demonstrate its ability to deliver electrons with a very narrow energy distribution.

\end{abstract}


\maketitle

The ability to control current flow down to the single electron level in mesoscopic devices has triggered a vast activity on quantum metrological current sources in recent years \cite{Geerligs1990,Kouwenhoven1991,Pothier1992,Martinis1994,Ono2003,Blumenthal2007,Pekola2008,Kaestner2008,Siegle2010,Giazotto2011,Giblin2012,Roche2013,Connolly2013,Pekola2013,Yamahata2014,Rossi2014,Stein2015}. 
In a quantum current source, electrons are conveyed one by one across a mesoscopic conductor, which is achieved owing to Coulomb repulsion. Early device geometries have been relying on two or more Coulomb blockaded islands in series \cite{Pothier1992}. Among the most promising recent approaches are islands with tunable barriers in 2D electron gases \cite{Kouwenhoven1991,Blumenthal2007,Giblin2012, Stein2015} along with superconducting single electron transistors \cite{Pekola2008}. 
Beyond metrological applications, the development of on-demand sources of single electrons opens paramount perspectives in the field of quantum coherent electronics and electron optics \cite{Feve2007,McNeil2011,Hermelin2011,Freulon2015}.

The superconducting single electron transistor (SINIS) turnstile \cite{Pekola2008,Averin2008} takes advantage of the sharply defined energy gap in the density of states of superconductors, as an energy filter.  A small normal metallic region (N) is weakly coupled to two superconducting leads (S) through tunnel barriers. 
N has to be sufficiently small to present a charging energy $U$, which should be at least on the order of the superconducting gap of the leads, $\Delta$. Nevertheless, N displays densely distributed states, appearing as continuous at accessible temperatures. 
A finite island temperature then allows for an entire energy window $\sim k_BT$ in N of available states for tunneling, which leads to turnstile operation errors associated to double occupation of N or tunneling into the wrong lead \cite{Averin2008}.

\begin{figure}[t]
  	\includegraphics[width=1.00\columnwidth]{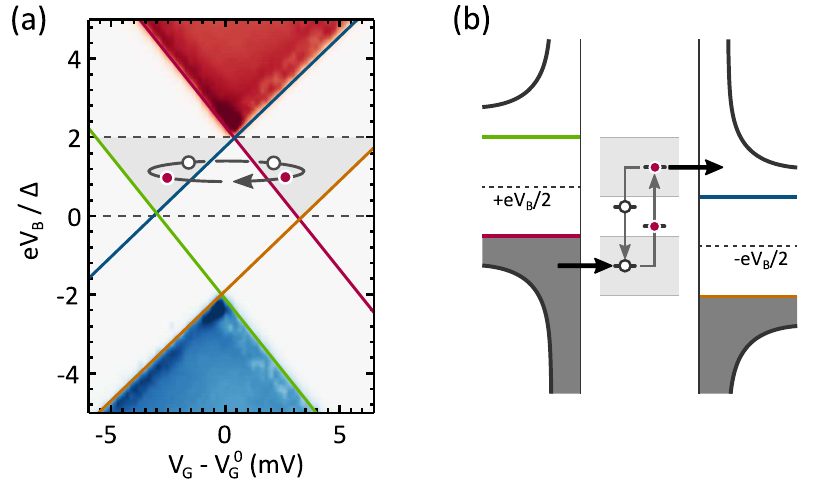}
	\caption{(a) Experimental current map of a superconductor - quantum dot hybrid device as function of gate and bias potential, in absence of periodic gate drive. Coloured solid lines correspond to the four superconducting gap edges as illustrated in (b). The device is operated as a single level turnstile when its state is modulated periodically around its $(n,n+1)$ charge degeneracy point. The on-state currents are $I_{+}=290 $ pA (red) and $I_{-}= -250$ pA (blue). (b) Energy diagram of the device with a small bias applied, illustrating electron tunneling events in and out the quantum dot. Grey areas indicate the amplitude range for solely forward tunneling, also seen in (a). Driving the turnstile with a square wave signal allows for tunneling at precisely determined energies.}
	\label{schematics}
\end{figure}

\begin{figure}
	\includegraphics[width=0.99\columnwidth]{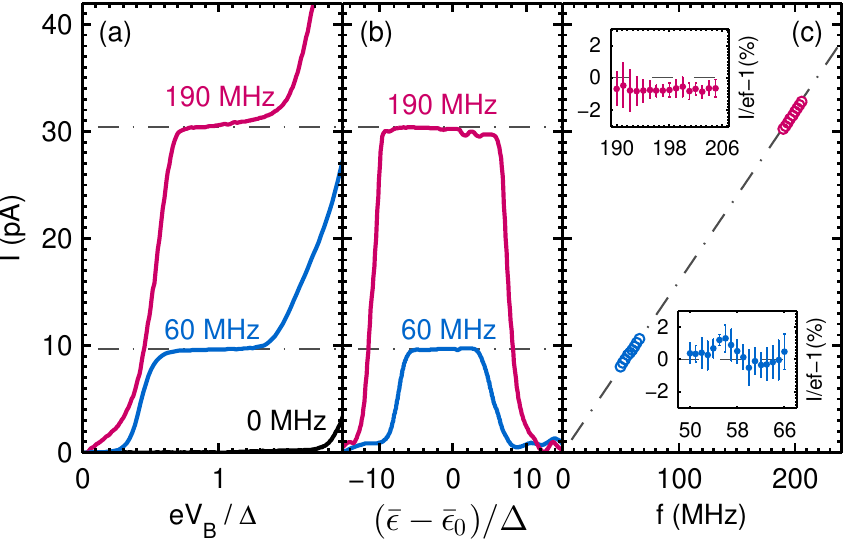}\hspace{4mm}
	\caption{(a) Current-bias traces measured near the charge degeneracy point. Characteristic plateaus appear with $I = ef$ (indicated by dashed lines) when applying a small modulation signal (magenta: $A_{\epsilon} \approx 0.64\Delta$, $f=190$ MHz, device $\mathcal  S$, blue: $A_{\epsilon} \approx 0.62\Delta$, $f=60$ MHz, $\mathcal A$) to the gate. The black trace shows the current response with no ac gate drive (device $\mathcal A$). (b) Current-gate traces measured for $A_{\epsilon} \approx \Delta$ and the same frequencies as in (a), at $V_B = \frac{3}{2}\Delta/e$ (magenta) and $V_B = \Delta/e$ (blue). (c) Current at the inflection point of the plateaus shown in (a) as function of operation signal frequency. The insets highlight deviations of the normalized current $I/ef$ from 1 in both the low and high frequency ranges.}
	\label{turnstile}
\end{figure} 

In this Letter, we demonstrate the first realization of a source of quantized current based on a single quantum energy level. The physical operation principle is similar to the SINIS turnstile, with the important difference that electrons are here carried by a single energy level of a quantum dot (Q). After demonstrating the expected principal turnstile operation characteristics, we focus on novel electronic transport features of the SQS turnstile. In particular, we show that 
tunneling can be tuned to occur at a precisely determined energy. We theoretically compare the dominant turnstile error processes in the SQS and SINIS devices respectively, concluding on the lower sensitivity of the former to out-of-equilibrium quasiparticles.

The fabrication of the SQS junctions, described in \cite{vanZanten15a}, relies on the {\it in situ} creation of a nanometer-sized fracture in superconducting constrictions. We perform controlled electromigration of on-chip metallic constrictions \cite{Park99}, which is a proven technique for connecting single molecules \cite{Park00} and gold nano-particles \cite{Kuemmeth08}. Randomly dispersed gold nanoparticles of about 5 nm diameter can occasionally bridge the nanometric fractures, providing thereby well-defined quantum dot junctions. By using superconducting aluminum electrodes, SQS junctions can be obtained \cite{Winkelmann09,vanZanten15a}. Because higher order processes are detrimental to turnstile operation accuracy, we restrict ourselves to rather weakly coupled devices.

The relevant device parameters of the quantum dot junction at the heart of the single-level turnstile are its charging energy $U$, the quantum dot orbital level spacing $\delta E$, the tunnel couplings $\gamma$ and the capacitances $C$ to the three terminals source, drain and gate, which we denote by indices $S$, $D$ and $G$ respectively in the remainder. 
All these can be determined from transport data in static conditions, that is, measuring the current $I$ as a function of the applied bias voltage $V_B$ and gate voltage $V_G$. The $I(V_B,V_G)$ maps show typical Coulomb blockade behavior in which only a single or at most a few charge degeneracy points (Fig. \ref{schematics}a) are accessible in the available gate voltage range. 
We find charging energies $U>50$ meV and orbital energy level spacings $\delta E$ on the order of 1 meV or higher.
Because $\delta E \gg k_BT$, the thermal population beyond the ground state is vanishingly small and electron transport occurs uniquely through a single orbital quantum level \cite{Ralph95}. Our study focusses on two devices with quite different tunnel couplings: $\mathcal S$ has rather symmetric tunnel couplings ($\gamma_S=2.1$ $\mu$eV, $\gamma_D=1.4$ $\mu$eV), while $\mathcal A$ is strongly asymmetric ($\gamma_S=5.2$ $\mu$eV, $\gamma_D=0.4$ $\mu$eV). We set $\hbar=1$ in the remainder, thereby identifying tunneling rates and energies. The determination of all dc transport characteristics of both devices was described in detail in \cite{vanZanten15a}.

Superconductivity in the leads provides a hard energy filter for tunneling.
The absence of quasi-particle states at energies $| E | < \Delta\approx$ 260 $\mu$eV  in the  leads results in a suppression of conductance for $|V_B |< 2\Delta/e$ at any gate voltage, as is seen in Fig. \ref{schematics}a. 
For turnstile operation, a small constant bias $0<|V_B| < 2\Delta/e$ is applied and a periodic modulation signal with frequency $f$ and variable amplitude  is added to the static gate potential. 
The energy difference between the $n+1$ and $n$ electron occupation numbers in the quantum dot, $\epsilon(t)$, varies between  ${\bar \epsilon} \pm A_{\epsilon}$, where $\bar \epsilon$ is controlled by the static voltages $V_G$ and $V_B$. 
A single electron can tunnel into the quantum dot  as soon as $\epsilon(t)$ faces the occupied states of the contact with the higher chemical potential (Fig. \ref{schematics}b; right grey triangle in Fig. \ref{schematics}a). By raising $\epsilon(t)$ via the back gate to face the empty states above the upper superconducting gap edge in the opposite lead (left grey triangle in Fig. \ref{schematics}a), the level is emptied to that lead. By operating the gate voltage cyclically, one electron is conveyed by cycle from the higher chemical potential lead to the other, giving rise to a dc current $I=ef$.

The combination of both above tunneling processes, in and out of the quantum dot, corresponds to the desired operation mode of the turnstile and will be named {\it forward} tunneling in the remainder. As can be seen in Fig. \ref{schematics}, forward tunneling requires the amplitude $ A_{\epsilon} $ of the modulation of $\epsilon (t)$  to verify $A_{\epsilon} > \Delta - e|V_B|/2$. On the other hand, a too large modulation amplitude $A_{\epsilon} > \Delta + e|V_B|/2$ will eventually allow for tunneling into/from the opposite lead. Such {\it backtunneling} processes are detrimental to current quantization, and their signature will be discussed later on.

 Throughout this work, a square wave signal, with a rise time  $\tau \approx$ 1.6 ns associated to the finite bandwidth of the generator, is used for modulating $\epsilon(t)$. 
The experimental dc current $I(V_B)$ measured for a square wave $\epsilon(t)$ with amplitude $A_{\epsilon}$ around ${\bar \epsilon}={\bar \epsilon}_0\equiv(\mu_S + \mu_D)/2$  is shown in Fig. \ref{turnstile}a. Here $\mu_{S,D}$ are the leads chemical potentials, such that $\mu_{S}-\mu_{D}=eV_B$. Above the threshold voltage for forward tunneling, $V_B^{fw} = \pm 2(\Delta-A_{\epsilon})/e$, a broad current plateau at $I=ef$ develops. Turnstile operation is only effective for a restricted range of values of $\bar \epsilon$ (Fig. \ref{turnstile}b). The value of the turnstile current, determined at the inflection point, follows the predicted linear dependence on frequency (Fig. \ref{turnstile}c), with a standard deviation of about 1 \%, to which adds a systematic deficit of about 0.7 \% at higher frequencies.
The plateaus show a small residual slope at all frequencies. 
This feature has instrumental origin, which is discussed in the Supplemental Material file.

\begin{figure}[t]
	\includegraphics[width=1.00\columnwidth]{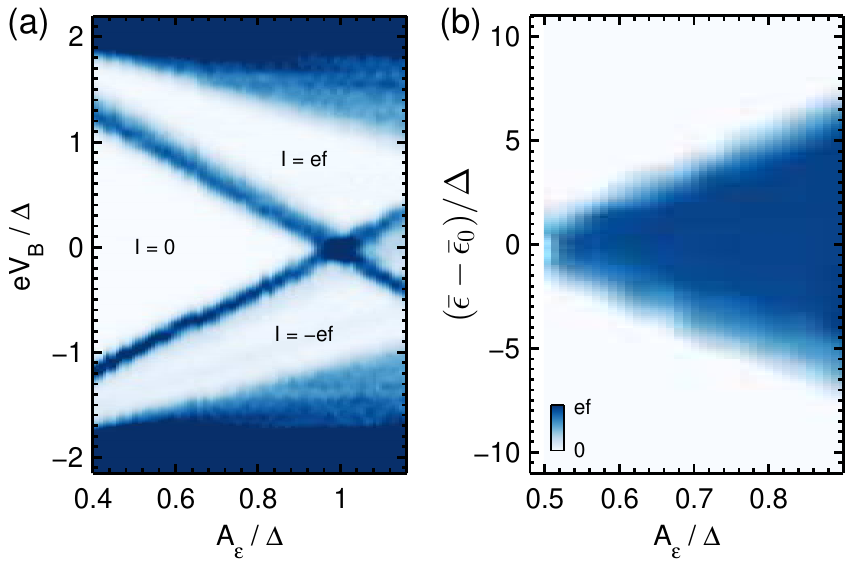}
	\caption{(a) Colormap of $\partial I/ \partial V_B$ of device $\mathcal S$ as a function of bias and gate modulation amplitude ($f = 56$ MHz, ${\bar \epsilon}={\bar \epsilon}_0$). Narrow blue regions, corresponding to rapid increase in current, separate areas of voltage independent current (white), with values  $I=0$ and $I=\pm ef$. (b) Colormap of turnstile current of $\mathcal S$ as a function of static gate offset from degeneracy point and gate modulation amplitude ($f = 60$ MHz, $V_B = 1.5\Delta/e$).}
	\label{backtunneling}
\end{figure}

At charge degeneracy, the thresholds for the onset of both forward and backtunneling can be seen as the narrow blue stripes  in Fig. \ref{backtunneling}a. Both threshold conditions cross at $V_B=0$ when $A_{\epsilon}=\Delta$. Whereas the frequency dependent transmission of the ac gate signal to the device is not precisely known, this crossing is used to calibrate $A_{\epsilon}$. The bright color identifies regions of voltage independent current, corresponding to $I=0$ and $I=\pm ef$ respectively.

When $\bar \epsilon$ is slightly detuned from  ${\bar \epsilon}_0$ by the static gate potential, the onset of forward tunneling is linearly shifted towards larger $A_{\epsilon}$ (Fig. \ref{schematics}b). Note that turnstile operation requires two successive tunneling events to successfully occur. This is visible in Fig. \ref{backtunneling}b, where the current is shown as function of gate detuning and modulation amplitude. For larger amplitudes $A_{\epsilon}$, an increasing tolerance of the turnstile operation with respect to the proper tuning of ${\bar \epsilon}-{\bar \epsilon}_0$ develops.

\begin{figure}
	\includegraphics[width=1.00\columnwidth]{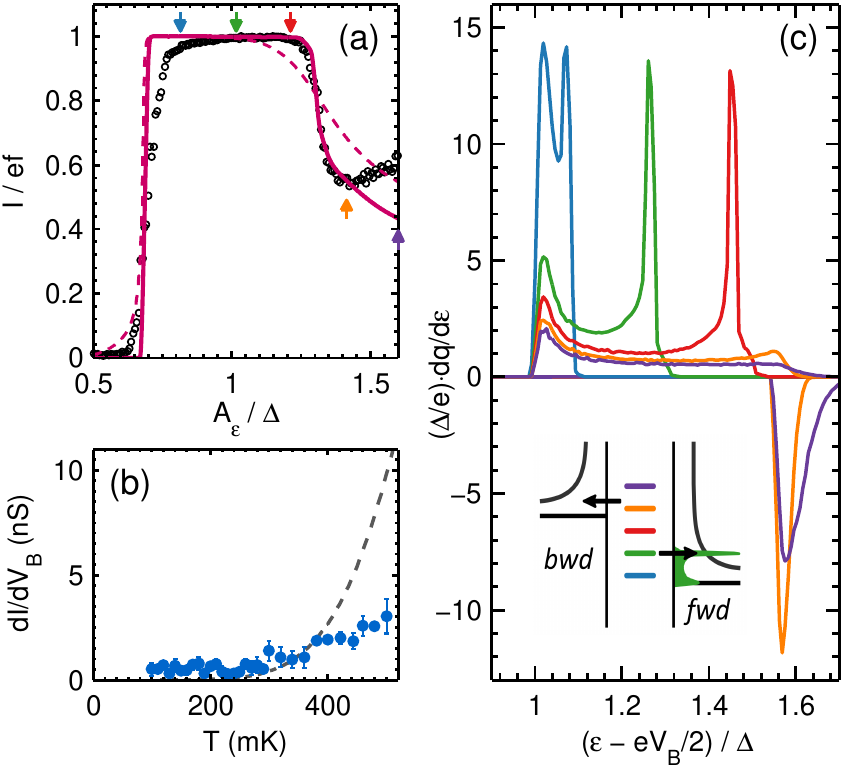}
	\caption{(a) Turnstile current  as a function of operation signal amplitude (device $\mathcal S$, $f = 56$ MHz, ${\bar \epsilon}={\bar \epsilon}_0$ and $eV_B =  0.7\Delta$. The sharp decrease in current indicates the sudden onset of back-tunneling. The continuous line is the numerical calculation for the SQS with all parameters determined by the device dc transport properties (see text). The dashed line is the analogous calculation for an SINIS device with parameters close to those of Ref. \cite{Knowles2012} (normal state resistance $R_N= 300$ k$\Omega$, $U=3.0 \Delta$, electron-phonon relaxation leading to quasi-equilibrium of electrons in N is included \cite{Giazotto2006}). The arrows indicate the values of $A_{\epsilon}$ used in (c) with corresponding colors. (b) Slope at inflection point of $I(V_b)$  on the turnstile plateaus, averaged over $A_{\epsilon}$, as a function of temperature (device $\mathcal A$). The dashed line is the calculation for the SINIS device, with parameters as in (a). (c) Calculation of the energy distribution of the delivered charge per cycle, for different gate drive amplitudes $A_\epsilon$, with parameters as in (a). The negative part of the panel displays the backtunneling contribution. The maximum relative excursion of the quantum dot level trajectory is represented in the inset for corresponding colors.}
	\label{temperature}
\end{figure}

Having evidenced electron turnstile operation, let us now identify the hallmarks of transport through a single quantum energy level. 
In SINIS turnstiles, backtunneling can be occasioned by electrons from the high energy tail of the thermal energy distribution in N. The probability of backtunneling increases thus steadily and smoothly as $A_{\epsilon}$ is cranked up \cite{Kemppinen2009}. Conversely, in a SQS turnstile backtunneling sets in abruptly, when  the threshold $A_{\epsilon} = \Delta + |V_B|/2e$ is exceeded. This is seen in Fig. \ref{temperature}a, where at high enough modulation amplitudes, the current starts dropping suddenly from $ef$. 
We numerically model the turnstile current dependence on $A_{\epsilon}$, both for the SINIS and the SQS turnstile, by solving the time-dependent rate equations using the measured output of the ac signal generator. In the case of the SQS, the instantaneous tunneling rates to each lead are found from the retarded Green's function's pole \cite{LevyYeyati1997,Kang98,vanZanten15a}, that is, beyond Fermi's golden rule. This is particularly important near the singularities in the superconducting density of states (see Supplemental Information file). The calculation (continuous line in Fig. \ref{temperature}a) nicely captures the abrupt decrease of the current as soon as the backtunneling threshold is met.  For comparison, in a SINIS device with parameters taken from the most precise devices presently studied \cite{Knowles2012, Nakamura2014}, the onset of backtunneling is markedly smoother (dashed line).

This particulary sharp onset of backtunneling is all the more pronounced if the rise time of $\epsilon(t)$ is short, or more precisely, if the time available for forward tunneling {\it only} is brief. 
If $\epsilon$ is raised to the backtunneling threshold within $\tau \ll \gamma^{-1}$, the probability of backtunneling may actually exceed that of forward tunneling. This means that  a current inversion, of magnitude up to $ef$, might eventually be produced with proper parameter combinations. This could however not be observed in our experiment, because the square wave rise time is of the same order of magnitude as the inverse tunneling rate ($\tau\sim  \gamma^{-1}$). 

As to highlight the energy selectivity of the tunneling process, we calculate the energy resolved transferred charge $dq/d\epsilon$ over a half-period of an ac gate cycle, using the assumptions and parameters of the calculation in Fig. \ref{temperature}a. The results are shown for different values of $A_{\epsilon}$ and for both forward and backward processes in Fig. \ref{temperature}c. While a certain fraction of forward tunneling occurs near the superconducting gap edge (where the lead's density of states is largest), good energy selectivity of the tunneling is achieved for sufficiently large values of $A_{\epsilon}$. The accuracy of the energy selectivity is ultimately limited by the tunnel coupling, but in the present experiment it is dominated by deviations of the ac drive signal from a perfect square wave. For even larger $A_{\epsilon}$, backtunneling is possible, which we represent using negative values of $dq/d\epsilon$. 

We now move to the discussion of possible error processes of the SQS turnstile. 
One obvious source of error in a turnstile is the missed tunneling event. As the tunneling rate is finite, tunneling may be missed during the corresponding half-period, leading to $I<ef$. For a single-level quantum dot, the Fermi golden rule tunneling rate for each lead ($\alpha=S,D$) can be written as $\Gamma_\alpha=(2)\,\gamma_\alpha \, n_s(\epsilon(t)\pm eV_b/2)$, where $n_s(E)$ is the normalized quasi-particle density of states in the superconducting leads. The factor of~$2$ takes into account the possibility of tunneling for two spin projections, and is present only for tunneling at one of the leads. For a symmetric square wave modulation of $\epsilon(t)$, the probability of missed tunneling at one of the leads can be roughly estimated as $e^{-\Gamma_\alpha{t}_{\rm{eff}}}$. Here, the effective time available for tunneling $t_{\rm{eff}}\approx1/(2f) -\tau$, takes into account the signal rise time. At frequencies around 200~MHz, this estimate gives a current deficit of 0.8 \% for the device parameters of sample $\mathcal S$, which agrees well with the experimental value of about 0.7 \% (Fig. \ref{turnstile}c inset).

In turnstile operation with a normal metal island and at finite temperature, an electronic population of magnitude $ \exp (-\Delta/k_BT)$ has sufficiently high energy for backtunneling. In aluminum-based devices, the associated error is rapidly dominant in SINIS turnstiles above about 300 mK \cite{Nakamura2014}. 
An expected hallmark of energy quantization in the turnstile operation should be a rather marked temperature insensitivity as long as $\delta E \gg k_BT$ and Pauli blocking of states in the leads can be neglected. We have followed the turnstile operation of device $\mathcal S$ as a function of temperature up to 0.5 K and we indeed observe the turnstile plateau to subsist through the entire temperature range, with only a rather moderate increase in error rate. We quantify the error by the $I=ef$ plateau slope $d I/d V_B$. As seen in Fig. \ref{temperature}b, this slope shows only little dependence on temperature. For comparison, the calculation of the same quantity for a SINIS turnstile with parameters $R_N= 300$ k$\Omega$, $U=3.0 \Delta$ is also plotted, showing a rapid divergence above $\sim$300 mK. While thermal errors are negligible only in the low mK range in most reported turnstiles, the SQS device can operate up to relatively high temperatures without suffering from thermal tunneling.

We now discuss a series of possible error processes for both the SQS and SINIS devices, up to third order in $\gamma/\Delta$.
For simplicity, we take $|eV_B|$, $A_\epsilon$ and $\Delta$ to be all of the same order, as is the case in usual turnstile operation conditions.  For the SINIS turnstile, the level spacing $\delta E\ll\Delta$ in N is small and any of about $\sim\Delta/\delta E$ electrons can tunnel. Writing $\gamma$ the tunnel coupling of an individual orbital level, the escape rate from N is $\Gamma \sim{g}\Delta$, where $g\sim\gamma/\delta E\ll{1}$~is the dimensionless conductance of the tunnel junctions, in units of the conductance quantum. Because $k_BT\ll U,\Delta$, real processes involving more than the two accessible charge states are suppressed, which also forbids Andreev reflection. Detailed derivations of the results given below are provided in the Supplemental Material file.

An important source of errors in superconducting turnstiles is related to the presence of non-equilibrium quasiparticles in the leads, with concentration $x_\mathrm{qp}=n_\mathrm{qp}/(2\nu\Delta)$, where $n_\mathrm{qp}$ is the quasiparticle density in the lead and $\nu$~is the density of states (per spin projection) at the Fermi level in the lead in normal state. Such quasiparticles can accumulate as a consequence of noise and, in particular, the turnstile operation itself and are well known to be very difficult to evacuate \cite{Knowles2012,Saira2012}. Using the diffusion model described in \cite{Knowles2012} we estimate a turnstile operation-induced non-equilibrium quasi-particle density on the order of 10 $\mu$m$^{-3}$ near the SQS junction, yielding $x_\mathrm{qp}\approx2\times10^{-6}$.
One first process involving  these quasiparticles is the direct tunneling between one lead and the central island, with rate $\sim{x}_\mathrm{qp}\,g\,\Delta$ in the  SINIS turnstile. 
Crucially, this process, which is at present the main source of errors in SINIS turnstiles \cite{Knowles2012}, is exponentially suppressed for the SQS device, as there are no available states in the quantum dot that allow to conserve energy. 

Beyond single particle processes,  a quasiparticle can be transferred from one lead to the other by cotunneling, that is, via an island level as a virtual intermediate state. Such second order  in $\gamma$ processes are allowed in both devices, with rates $\sim{x}_\mathrm{qp}g^2\Delta$ for the SINIS and $\sim{x}_\mathrm{qp}\gamma^2/ \Delta$ for the SQS, respectively. 
Eventually, the ultimate limitation to accuracy of superconducting turnstiles arises from third-order Cooper-pair-electron (CPE) cotunneling~\cite{Kang1999,Averin2008}. This process is effective even in the absence of quasiparticles
and its rate can be estimated as $\sim{g}^3\Delta $ and $\sim\gamma^3/\Delta^2$ for the SINIS and SQS devices, respectively.
To compare the amplitude of all above mechanisms in the SINIS and SQS devices, let us assume both to display the same forward tunneling rate, that is, $g\Delta\sim\gamma$.  The quasiparticle and the CPE cotunneling rates are then comparable in both, as the larger number of states in N is compensated by a lower tunnel coupling strength per state. The higher order processes discussed above all lead to an excess current with respect to $ef$. 

To conclude, a metallic quantum dot embedded between superconducting leads allows for turnstile operation in which the charges are conveyed by a single orbital quantum level. As a consequence, tunneling occurs precisely at the quantum level energy. Under realistic assumptions, the SQS turnstile can serve as a monochromatic on-demand single electron source. As a next step, one can explore the possibility of spin polarized turnstile operation by Zeeman splitting the orbital quantum level using a moderate magnetic field.

This work was funded by the European Union Seventh Framework Programme INFERNOS (FP7/2007-2013) under grant agreement no. 308850 and partially (JPP and IMK)  by the Nanosciences Foundation, foundation under the aegis of the Joseph Fourier University Foundation. Samples were fabricated at the Nanofab facility at Institut N\'eel. We also thank F. Hekking, X. Waintal, and B. Sac\'ep\'e for discussions and C. Hoarau, A. Nabet, F. Balestro  and A. de Cecco for help with the experiments. We are indebted to E. Bonet, C. Thirion and W. Wernsdorfer for developing and sharing NanoQT.

\bibliography{./Literature}

\pagebreak
\widetext
\begin{center}
\textbf{\large Supplemental Materials: Single Quantum Level Electron Turnstile}
\end{center}
\setcounter{equation}{0}
\setcounter{figure}{0}
\setcounter{table}{0}
\setcounter{page}{1}
\makeatletter
\renewcommand{\theequation}{S\arabic{equation}}
\renewcommand{\thefigure}{S\arabic{figure}}
\renewcommand{\bibnumfmt}[1]{[S#1]}
\renewcommand{\citenumfont}[1]{S#1}

\newcommand{\ep}{\epsilon}
\newcommand{\vep}{\varepsilon}
\newcommand{\hc}{{\rm \;h.\,c.\;}}
\newcommand{\sign}{\mathop{\mathrm{sign}}\nolimits}
\renewcommand{\Im}{\mathop{\mathrm{Im}}\nolimits}
\renewcommand{\Re}{\mathop{\mathrm{Re}}\nolimits}

\hypersetup{urlcolor=blue}

This document provides supplemental material to the manuscript {\it Single quantum level electron turnstile}, discussing in particular  some instrumental parasitic effects in the turnstile operation and theoretical aspects of the Green's functions calculations of both the tunneling rates and error processes.

\section{Parasitic $V_B$ oscillations}
\begin{figure}[b]
	\centering
	\includegraphics[width=10.0cm]{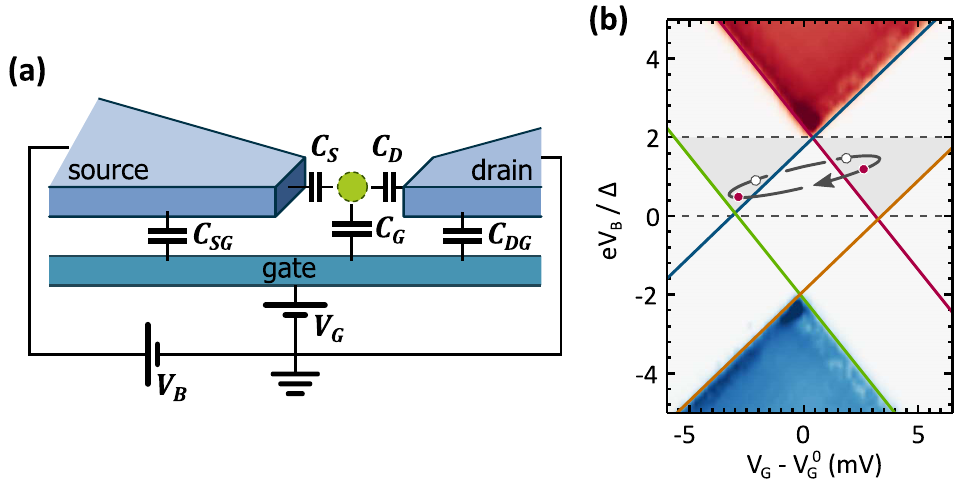}
	\includegraphics[width=7cm]{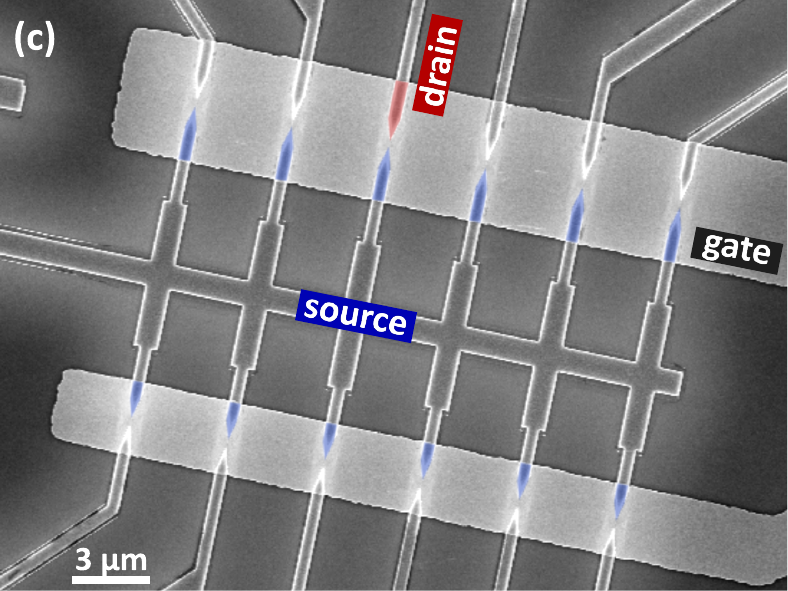}
	\caption{a) Schematic of the SQS device illustrating all relevant capacitances. b) Replication of $I(V_B,V_G)$ map shown in Fig. 1a of the main article, illustrating a tilted quantum dot state trajectory due to the effective gate-bias crosstalk. (c) Colored large view scanning electron micrograph of the chip, displaying 12 electromigration junctions. The bright grey region is the local backgate, isolated from the source and drain leads by an 8 nm thick aluminum oxide. The source is common to all 12 junctions. The red and blue regions highlight the ensuing asymmetry in capacitive coupling to the gate of the source and drain contacts of a given electromigration junction.}
	\label{fig_2}
\end{figure}

Under time-invariant bias conditions one can neglect the effect of the capacitance between the leads and the gate electrode. However, with the application of a time-dependent potential $V_G(t)$ to the gate, the electronic potentials of the leads near the junction can become affected through these capacitances, $C_{SG}$ and $C_{DG}$ respectively. If $C_{DG} \approx C_{SG}$, the voltage drop over the junction is only little affected \cite{S_Maisi2009}. In the present experiment however, we can estimate from the device geometry that $C_{SG}$ and $C_{DG}$ differ by more than one order of magnitude (Fig. \ref{fig_2}c). This rather large asymmetry results in adding a small oscillating component to the bias. Equivalently, the trajectory of $\epsilon(t)$ in the $(V_G,V_B)$-plane can be pictured as a slightly tilted ellipse, as schematically shown in Fig. \ref{fig_2}b.

\begin{figure}[t]
	\centering
	\includegraphics[width=17cm]{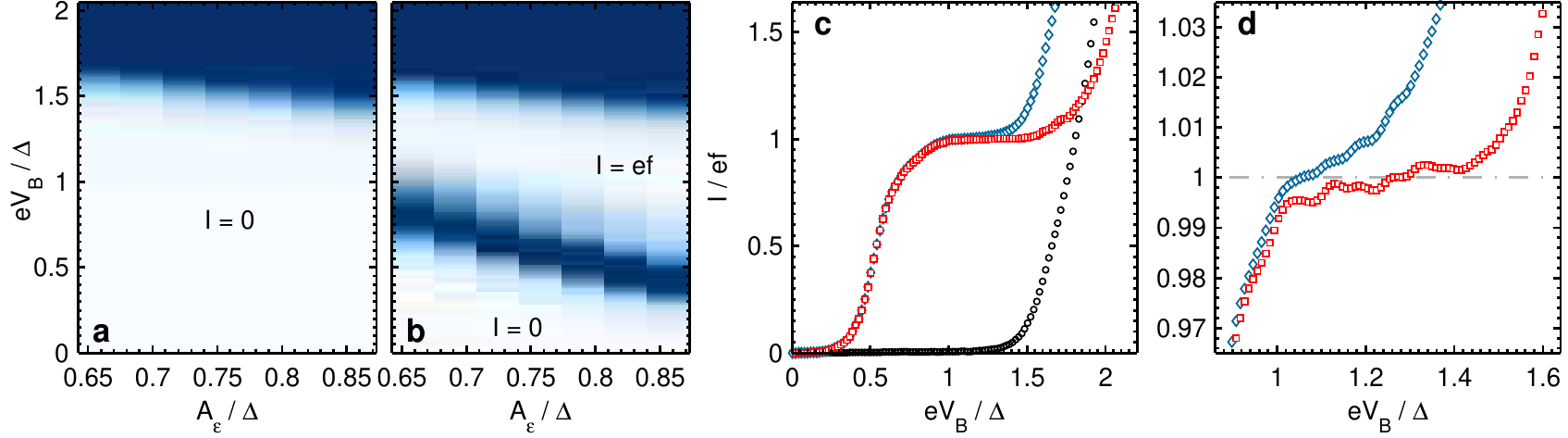}
	\caption{Differential conductance maps as a function of applied bias voltage $V_B$ and gate modulation amplitude $A_\ep$ measured while the static gate voltage $V_G$ is tuned far from (a) and right at (b) the charge degeneracy point (device A, $f=60$ MHz, color code as in Fig. 3a of main manuscript). (c) Turnstile current at $A_\ep = 0.85\Delta$ as a function of the static bias voltage $V_B$ while the static gate voltage $V_G$ is tuned far from ($\circ$) and right at ($\diamond$) the charge degeneracy point.  The difference between both traces is plotted using squares. (d) Zoom on $I=e\,f$ region. The subtraction reveals  a much broader plateau with a significantly reduced slope.}
	\label{fig_3}
\end{figure}

Such a tilt does not necessarily affect the turnstile current as long as the time dependent $V_B(t)$ remains below $2\Delta/e$. Above this threshold, direct SIS tunnel events and co-tunneling events are no longer strictly suppressed by energy conservation. Consequently, this can give rise to additional current during a finite time window per signal period, which increases with the static $V_B$ and $A_\ep$. Fig. \ref{fig_3} compares the response of the turnstile  as a function of ($V_B,A_\ep$), depending on  whether the static gate potential $\bar{V}_G$ is tuned far away from the charge degeneracy point (a) or at charge degeneracy (b). As expected, in (b) turnstile operation is observed for bias voltages beyond the blue lines, just as in Fig. 3a of the main manuscript. As expected again, no turnstile current is observed in (a). Nevertheless, both panels display a small yet identical slope of the threshold in $V_B$ above which current quantization is lost, as a function of $A_\ep$.

From the above observations, we conclude that the experimentally observed slope of the $I=e\, f$ plateaus has its origin essentially in the gate-bias crosstalk discussed above. 
 As to more quantitatively demonstrate the contribution of this effective cross-talk, one can take the difference between the measurements displayed in Fig. \ref{fig_3}(a,b). Figure \ref{fig_3}c shows $I(V_B)$ traces of the left (right) panel in black (blue) respectively, at a given $A_\ep$, as well as their difference in red. A zoom on the $I=e\,f$ plateau (Fig. \ref{fig_3}d) evidences the striking improvement of the turnstile accuracy when corrected for the gate-bias cross-talk. Next generation experiments will take care to minimize this parasitic effect by making the gate-to-lead capacitances more symmetric.

\section{Electron tunneling rate beyond Fermi's golden rule}
In many cases, electron tunneling rate from the dot into the
electrodes is found from Fermi's golden rule.
Coupling of the dot level
$\epsilon$ to the electrodes controlled by an external gate voltage $V_G$ can be conveniently described by the
(retarded) self-energy $\Sigma(E)$, $E$ is a quasiparticle energy. The standard perturbative
Fermi's golden rule approach corresponds to approximating the tunneling
rate $\Gamma$ by $\Gamma\approx-2\Im\Sigma(\epsilon)$.
This approximation is valid provided that the energy dependence
of $\Sigma(E)$ near $E=\epsilon$ is smooth on the scale of
$\Gamma$.

For a single superconducting electrode, the self-energy
\begin{equation}
\Sigma(E)=-\frac{\gamma E}{\sqrt{\Delta^2-E^2}},
\end{equation}
where $\gamma$ is half the decay rate for the electrode in the
normal state. Clearly, Fermi's golden rule breaks down near the
BCS singularities $E\to\pm\Delta$. To go beyond Fermi's golden rule,
one has to solve the Green's function pole equation,
\begin{equation}\label{poleequation=}
E-\epsilon-\Sigma(E)=0,
\end{equation}
in the lower complex half-plane of~$E$, as the retarded Green's
function $G_d^R(E)=[E+i0-\epsilon-\Sigma(E)]^{-1}$, as well as the
retarded self-energy $\Sigma(E)$ must be analytical in the upper
half-plane. Having found the pole at some $E=E_*$, one can
associate its imaginary part with the decay rate,
$\Gamma=-2\Im E_*=-2\Im \Sigma(E_*)$.

At the backtunneling onset we must take into account coupling to
both electrodes, so the self-energy has two contributions,
\begin{equation}\label{SigmaLSigmaR=}
\Sigma(E)=\Sigma_S(E)+\Sigma_D(E) \\
=  -\frac{\gamma_S(E-e V_B/2)}{\sqrt{\Delta^2-(E-e V_B/2)^2}}
-\frac{\gamma_D(E+e V_B/2)}{\sqrt{\Delta^2-(E+e V_B/2)^2}}.
\end{equation}
We focus on the electron tunneling, $ \Re E>0$. The hole tunneling
can be treated symmetrically, with $\gamma_S,\gamma_D$ multiplied
by~2 to account for the spin degeneracy. As we use the convention
with $e>0$ (the electron charge thus being $-e$),
$\Sigma_D$~describes the forward tunneling and $\Sigma_S$~the
backward one.

The decay rate can still be found from the imaginary part of the
Green's function pole~$E_*$, obtained by solving
Eq.~(\ref{poleequation=}). However, it gives the total decay
rate, which should still be separated into partial contributions
from the two electrodes. Such separation is straightforward when
$\Sigma_D(E)$ is a smooth function of $E$ near
$E=\Delta+e V_B/2$. Then, one can approximate
$\Sigma_D(E)\approx\Sigma_D(E_*)\approx-i\Gamma_D/2$,
where $\Gamma_D$ is the decay rate to the drain electrode.
The pole $E_*$ is shifted with respect to $\epsilon=\Delta+e V_B/2$ along the
real axis by an amount $\sim(\gamma_S^2\Delta)^{1/3}$, as
a result of level repulsion between the dot level and the BCS
singularity in the density of states in the source electrode. As
$(\gamma_S^2\Delta)^{1/3}\gg\gamma_S\sim\gamma_D$,
this repulsion is stronger than the level broadening by the
drain electrode $\Gamma_D\sim\gamma_D$. The criterion for the smoothness of
$\Sigma_D(E)$ is
\begin{equation}
|\Sigma_D(E)-\Sigma_D(E_*)|\approx(\gamma_S^2\Delta)^{1/3}
\left.\frac{\partial|\Sigma_D(E)|}{\partial E}
\right|_{E=\Delta+e V_B/2}\ll\gamma_D,
\end{equation}
which results in $e V_B\gg[(\gamma_S)^4\Delta^5]^{1/9}$.
In the opposite case of small~$V_B$, the two BCS singularities
overlap and we see no simple way to separate the two rates.

The above reasoning assumed $\epsilon$ to be time-independent. As the
typical rise time of $\epsilon$ in the experiment is $\tau\approx1.6$~ns, and
$-\Im\Sigma_D\gtrsim\gamma_D$, where $\gamma_D$ is several $\mu\mbox{e V}$, this assumption
indeed holds at the backtunneling onset.

To summarize, we calculate the tunneling rates at the backtunneling
onset as follows. First, we find the root $E_*$ of the pole
equation~(\ref{poleequation=}) with the
self-energy~(\ref{SigmaLSigmaR=}). Then, the tunneling rate to the
drain electrode is evaluated as $\Gamma_D=-2\Im\Sigma_D(E_*)$,
and the one to the source electrode is
$\Gamma_S=-2\Im \Sigma(E_*)-\Gamma_D$.

\section{Theoretical discussion of error processes in superconducting turnstiles}

Here we estimate the rates of various parasitic processes for multi-level and single-level dots, following Ref.~\cite{S_Averin2008}. All rates are obtained using the perturbative Fermi's golden rule.

\subsection{The model for the quantum dot and superconducting electrodes}

Consider a quantum dot coupled to two superconducting leads. Each lead
$\alpha=S,D$ is described by the BCS Hamiltonian
\begin{eqnarray}
&&\hat{H}_\alpha=\sum_{k,\sigma}\,\vep_k
\hat{a}^\dagger_{\alpha{k}\sigma}\hat{a}_{\alpha{k}\sigma},
\quad\vep_k=\sqrt{\Delta^2+\xi_k^2},\\
&&\hat{a}_{\alpha{k}\uparrow}=
\hat{c}_{\alpha{k}\uparrow}\sin\frac{\eta_k}{2}
+\hat{c}_{\alpha{k}\downarrow}^\dagger\cos\frac{\eta_k}{2},\quad
\hat{a}_{\alpha{k}\downarrow}=
\hat{c}_{\alpha{k}\downarrow}\sin\frac{\eta_k}{2}
-\hat{c}_{\alpha{k}\uparrow}^\dagger\cos\frac{\eta_k}{2},\\
&&\hat{c}_{\alpha{k}\uparrow}=
\hat{a}_{\alpha{k}\uparrow}\sin\frac{\eta_k}{2}
-\hat{a}_{\alpha{k}\downarrow}^\dagger\cos\frac{\eta_k}{2},\quad
\hat{c}_{\alpha{k}\downarrow}=
\hat{a}_{\alpha{k}\downarrow}\sin\frac{\eta_k}{2}
+\hat{a}_{\alpha{k}\uparrow}^\dagger\cos\frac{\eta_k}{2},\\
&&\eta_k=\frac\pi{2}+\arctan\frac{\xi_k}\Delta,\quad
\sin\frac{\eta_k}{2}\equiv{u}_k,\quad\cos\frac{\eta_k}{2}\equiv
v_k=\sqrt{\frac{1-\xi_k/\vep_k}{2}},
\end{eqnarray}
where $k$~labels the orbital states in each electrode,
$\hat{c}^\dagger,\hat{c}$ are creation and annihilation operators
for electrons, and $\hat{a}^\dagger,\hat{a}$ those for Bogolyubov
quasiparticles. $\xi_k$~is the electron energy counted from the
Fermi level in the absence of the bias, and $\vep_k$~is the
quasiparticle energy.

To describe the multi-level quantum dots, we introduce the orbital
quantum number~$n$ to label the orbital states in the dot. For a
single-level dot, we will just assume that $n$ can take just one
value.
The single-particle part of the dot Hamiltonian and its coupling to
the leads is written as
\begin{equation}
\hat{H}_\mathrm{dot+tun}=\sum_{n,\sigma}
E_n\hat{c}_{n\sigma}^\dagger\hat{c}_{n\sigma}+
\sum_{\alpha=S,D}\left(\hat{V}_\alpha+\hat{V}_\alpha^\dagger\right),
\quad
\hat{V}_\alpha=\sum_{k,n,\sigma}
W_{\alpha{k}n}\hat{c}^\dagger_{\alpha{k}\sigma}\hat{c}_{n\sigma}.
\end{equation}
The energy levels $E_n$ are assumed to have the mean level
spacing~$\delta E$.
The tunneling matrix elements can be modeled as~\cite{S_Aleiner1996}
$W_{\alpha{k}n}=W_\alpha\,\phi^*_{\alpha{k}}(\vec{r}_\alpha)\,
\phi_n(\vec{r}_\alpha)$, where the wave functions $\phi_n(\vec{r})$
of a multilevel dot are treated as Gaussain random variables with
average $\overline{\phi_n(\vec{r}_\alpha)}=0$,
$\overline{\phi_n(\vec{r}_\alpha)\,\phi_{n'}(\vec{r}_{\alpha'})}
=\delta_{\alpha\alpha'}\delta_{nn'}$.
The matrix elements determine the decay rate of each dot level
and the conductance of each tunnel junction in the normal state,
\begin{equation}
2\gamma_{n\alpha}=2\pi|W_\alpha|^2\nu_\alpha
|\phi_n(\vec{r}_\alpha)|^2,\quad
G_\alpha=2{e}^2\,\frac{2\pi|W_\alpha|^2\nu_\alpha}{\delta E}
=\frac{2e^2}{2\pi}\,g_\alpha,
\end{equation}
where $\delta E$ is the mean level spacing on the dot, and
\begin{equation}
\nu_\alpha=\sum_k|\phi_{\alpha{k}}(\vec{r}_\alpha)|^2
\delta(\xi_{\alpha{k}})
\end{equation}
is the local density of states per spin in the $\alpha$th
electrode in the normal state, assumed to be energy-independent.

The electrostatic part of the energy is obtained by assuming capacitive
coupling to the source, drain, and gate electrodes via the capacitances
$C_S,C_D,C_G$. If voltages $V_S=-V/2$, $V_D=V/2$, and $V_G$ are applied
to the electrodes, the voltage on the dot as a function of its charge
$Q$ is given by
\begin{equation}
V_\mathrm{dot}(Q)=\frac{Q}{C_\Sigma}
+\frac{(C_S-C_D)V/2+C_GV_G}{C_\Sigma},\quad
C_\Sigma\equiv{C}_S+C_D+C_G.
\end{equation}
The electrostatic energy of the dot with $N$ electrons is then given by
\begin{equation}
E_N=U N(N-2N_G)+\frac{C_S-C_D}{2C_\Sigma}\,N e V_B,\quad
U=\frac{e^2}{2C_\Sigma},\quad N_G=-\frac{C_G V_G}e.
\end{equation}

\subsection{Sequential single-particle tunneling}

Let $p_N$ be the probability for the quantum dot to have $N$~electrons.
It satisfies the rate equation
\begin{equation}
\frac{dp_N}{dt}=\sum_{\alpha=S,D}
\left(-\Gamma^\alpha_{N\to{N}+1}p_N
-\Gamma^\alpha_{N\to{N}-1}p_N+\Gamma^\alpha_{N-1\to{N}}p_{N-1}
+\Gamma^\alpha_{N+1\to{N}}p_{N+1}\right),
\end{equation}
where the integrated transition probabilities are given by
(the factor of 2 from spin)
\begin{eqnarray}
&&\Gamma_{N+1\to{N}}^\alpha=2\sum_{n,k}|W_{\alpha{k}n}|^2f_n
\left[u_k^2(1-f_k^\alpha)\,2\pi\delta(E_{N+1}+E_n-E_N-e V_\alpha-\vep_k)
\right.+\nonumber\\
&&\hspace*{2.5cm} +\left.v_k^2f_k^\alpha\,
2\pi\delta(E_{N+1}+E_n+e V_\alpha+\vep_k-E_N-2e V_\alpha)
\right],\\
&&\Gamma_{N\to{N}+1}^\alpha=2\sum_{n,k}|W_{\alpha{k}n}|^2(1-f_n)
\left[u_k^2f_k^\alpha\,2\pi\delta(E_N+e V_\alpha+\vep_k-E_{N+1}-E_n)
\right.+ \nonumber\\
&&\hspace*{2.5cm}+\left.v_k^2(1-f_k^\alpha)\,2\pi\delta(E_N+2e V_\alpha
-E_{N+1}-E_n-\vep_k-e V_\alpha)\right].
\end{eqnarray}
Let us denote $U_\alpha=E_N-E_{N+1}+e V_\alpha$ and introduce
the distribution functions
\begin{equation}
f_n=f(E_n)=\frac{1}{e^{\beta E_n}+1},\quad
f_k^\alpha=f_\mathrm{qp}^\alpha(\vep_k)
=\frac{n_\mathrm{qp}^\alpha e^{-\beta^\alpha_\mathrm{qp}\vep_k}}%
{4\nu_\alpha\Delta\,K_1(\beta^\alpha_\mathrm{qp}\Delta)}\ll{1},
\end{equation}
where the quasiparticle density
$n_\mathrm{qp}^\alpha=2\sum_kf^\alpha_k|\phi_{\alpha{k}}(\vec{r}_\alpha)|^2$,
and the modified Bessel function
$K_1(z)=\int_0^\infty{e}^{-z\cosh\eta}\cosh\eta\,d\eta$.
Here we assumed that $f_k^\alpha$
depends only on $\vep_k$ and not on $\sign\xi_k$, thereby neglecting
the imbalance. Then we can extend it on the negative energies by
defining $f^\alpha_\mathrm{qp}(\vep<0)=1-f^\alpha_\mathrm{qp}(-\vep)$,
and write for an arbitrary function $\mathcal{F}(\vep)$
\begin{eqnarray}
&&W_\alpha^2\sum_k|\phi_{\alpha{k}}(\vec{r}_\alpha)|^2
\left[u^2_kf_k^\alpha\mathcal{F}(\vep_k)
+v_k^2(1-f_k^\alpha)\mathcal{F}(\vep_k)\right]=\nonumber\\
&&=W_\alpha^2\nu_\alpha\int{d}\xi_k\left\{\frac{\vep_k+\xi_k}{2\vep_k}\,
f^\alpha_\mathrm{qp}(\vep_k)\,\mathcal{F}(\vep_k)
+\frac{\vep_k-\xi_k}{2\vep_k}\,
[1-f^\alpha_\mathrm{qp}(\vep_k)]\,\mathcal{F}(-\vep_k)\right\}
=\nonumber\\
&&=\frac{\overline{\gamma}_\alpha}\pi\int\limits_{-\infty}^\infty {d}\vep\,
\frac{\theta(|\vep|-\Delta)|\vep|}{\sqrt{\vep^2-\Delta^2}}\,
f^\alpha_\mathrm{qp}(\vep)\,\mathcal{F}(\vep)
\equiv\frac{\overline{\gamma}_\alpha}\pi
\int\limits_{-\infty}^\infty\mathcal{N}_S(\vep)\,d\vep\,
f^\alpha_\mathrm{qp}(\vep)\,\mathcal{F}(\vep).
\end{eqnarray}
Then, for a multilevel dot we have
\begin{eqnarray}
&&\Gamma_{N+1\to{N}}^\alpha
=\frac{g_\alpha}\pi\int\limits_{-\infty}^\infty\mathcal{N}_S(\vep)\,d\vep\,
f(U_\alpha+\vep)\left[1-f_\mathrm{qp}^\alpha(\vep)\right],\\
&&\Gamma_{N\to{N}+1}^\alpha
=\frac{g_\alpha}\pi\int\limits_{-\infty}^\infty\mathcal{N}_S(\vep)\,d\vep\,
\left[1-f(U_\alpha+\vep)\right]f_\mathrm{qp}^\alpha(\vep).
\end{eqnarray}
For a single-level dot $f(U_\alpha+\vep)$ and $1-f(U_\alpha+\vep)$
should be replaced by $\delta E\cdot\,\delta(U_\alpha+\vep)$.

Let us now consider turnstile operation with $-V_S=V_D=V/2$,
$e V_B/\Delta\equiv\chi$, $0<\chi<2$.
On $N+1\to{N}$, electron can be ejected
into the drain electrode but not into the source one if
$\Delta-(-e V_B/2)<E_{N+1}-E_N<\Delta+(-e V_B/2)$.
On $N\to{N+1}$, electron can be injected
from the source electrode but not from the drain one if
$-\Delta-(-e V_B/2)<E_{N+1}-E_N<-\Delta+(-e V_B/2)$.
Thus we write $E_{N+1}-E_N=\pm(1-\chi/2+\eta)\Delta$ with $0<\eta<\chi$,
and for $N+1\to{N}$ we have
$U_D/\Delta=-(1+\eta)$, $U_S/\Delta=\chi-(1+\eta)$,
while for $N\to{N+1}$ we have
$U_D/\Delta=1+\eta-\chi$, $U_S/\Delta=1+\eta$.

Let us focus on the $N+1\to{N}$ stage at zero temperature. Then,
for a multilevel dot we have
\begin{eqnarray}
&&\Gamma_{N+1\to{N}}^D=\frac{g_D\Delta}\pi\sqrt{\eta(2+\eta)},\quad
\eta\gg\beta_\mathrm{qp}^D\Delta,\\
&&\Gamma_{N+1\to{N}}^S=\frac{g_S}\pi\,\frac{n_\mathrm{qp}^S}{4\nu_S}.
\end{eqnarray}
In addition, on the same stage, after the electron ejection,
there is a possibility of populating the dot again by injecting
a quasiparticle from $S$~electrode on a high level, with the same rate $\Gamma_{N\to{N+1}}^S=\Gamma_{N+1\to{N}}^S$.

For a single-level dot, both parasitic processes are absent,
as there are no filled or empty levels at the corresponding energies:
\begin{eqnarray}
&&\Gamma_{N+1\to{N}}^D=\frac{g_D\delta E}\pi\,\frac{1+\eta}{\sqrt{\eta(2+\eta)}}
\left[1-f_\mathrm{qp}^D(-U_D)\right],\\
&&\Gamma_{N+1\to{N}}^S=\Gamma_{N\to{N+1}}^S=0.
\end{eqnarray}

\subsection{Quasiparticle cotunneling}

To transfer an electron from electrode $\alpha$ to $\alpha'$,
we act with $\hat{c}^\dagger_{\alpha'k'\sigma'}\hat{c}_{n\sigma'}
\hat{c}^\dagger_{n'\sigma}\hat{c}_{\alpha{k}\sigma}$ if there are
$N$ electrons on the dot initially, and with
$\hat{c}^\dagger_{n'\sigma}\hat{c}_{\alpha{k}\sigma}
\hat{c}^\dagger_{\alpha'k'\sigma'}\hat{c}_{n\sigma'}$ if there
were $N+1$ electrons.
Starting from a given configuration of quasiparticles on the
electrodes, we can create or destroy a quasiparticle on each
electrode, so there are four possible final states which we
label by $s_\alpha=\pm$ and $s_{\alpha'}=\pm$. The inelastic
cotunneling rate is
\begin{eqnarray}
&&\Gamma^{\alpha\to\alpha'}_{N,\mathrm{in}}=4\sum_{k,k',n,n'}
\sum_{s_\alpha,s_{\alpha'}=\pm}
\left|\frac{W_{\alpha'k'n}W^*_{\alpha{k}n'}}%
{E_N+e V_\alpha+s_\alpha\vep_k-E_{N+1}-E_{n'}}\right|^2f_n(1-f_{n'})
\times\nonumber\\ &&\hspace*{3cm}{}\times
\frac{\vep_k+s_\alpha\xi_k}{2\vep_k}\,
\frac{1-s_\alpha(1-2f_k^\alpha)}{2}
\frac{\vep_{k'}+s_{\alpha'}\xi_{k'}}{2\vep_{k'}}\,
\frac{1+s_{\alpha'}(1-2f_{k'}^{\alpha'})}{2}
\times\nonumber\\ &&\hspace*{3cm}{}\times
2\pi\delta(e V_\alpha+s_\alpha\vep_k+E_n
-e V_{\alpha'}-s_{\alpha'}\vep_{k'}-E_{n'}),
\end{eqnarray}
where the factor 4 comes from spin summation.
For $N+1$ the only change is that the denominator becomes
$E_{N+1}+E_n-E_N-e V_{\alpha'}-s_{\alpha'}\vep_{k'}$, which
gives the same if the energy conservation is taken into account.
Thus, we can write
\begin{eqnarray}
&&\Gamma^{\alpha\to\alpha'}_{\mathrm{in}}=
\frac{g_\alpha{g}_{\alpha'}}{(2\pi^2)^2}
\int\limits_{-\infty}^\infty{d}E\,dE'\,
\mathcal{N}_S(\vep)\,d\vep\,\mathcal{N}_S(\vep')\,d\vep'\,
f(E)\,f^\alpha_\mathrm{qp}(\vep)[1-f(E')][1-f^{\alpha'}_\mathrm{qp}(\vep')]
\times{}\nonumber\\&&\hspace*{3cm}{}\times
\frac{2\pi\delta(U_\alpha+\vep-E'-U_{\alpha'}-\vep'+E)}%
{(U_\alpha+\vep-E')^2}=\nonumber\\
&&\qquad\mathop{=}\limits_{\beta\to\infty}
\frac{g_{\alpha}g_{\alpha'}}{2\pi^3}\int\limits_{-\infty}^\infty
\mathcal{N}_S(\vep)\,d\vep\,\mathcal{N}_S(\vep')\,d\vep'
f^\alpha_\mathrm{qp}(\vep)\left[1-f^{\alpha'}_\mathrm{qp}(\vep')\right]
\times{}\nonumber\\
&&\hspace*{3cm}{}\times
\left(\frac{1}{U_{\alpha'}+\vep'}-\frac{1}{U_{\alpha}+\vep}\right)
\theta(U_{\alpha}+\vep-U_{\alpha'}-\vep').
\end{eqnarray}
The cotunneling current is
$I_\mathrm{in}=e\Gamma^{S\to D}_\mathrm{in}
-e\Gamma^{D\to S}_\mathrm{in}$. If $\beta_\mathrm{qp}|e V_B|\gg{1}$,
we can neglect
$\Gamma^{D\to S}_\mathrm{in}\sim\Gamma^{S\to D}_\mathrm{in}
e^{-\beta_\mathrm{qp}|e|V}$.
\begin{eqnarray}
&&\Gamma^{S\to D}_\mathrm{in}\approx\frac{g_Sg_D}{2\pi^3}
\int\limits_\Delta^\infty{d}\vep\,\mathcal{N}_S(\vep)\,f_\mathrm{qp}^S(\vep)
\int\limits_\Delta^{\vep-e V_B}d\vep'\,\mathcal{N}_S(\vep')
\left(\frac{1}{U_D+\vep'}-\frac{1}{U_S+\vep}\right)+{}\nonumber\\
&&\quad\qquad{}+\frac{g_Sg_D}{2\pi^3}
\int\limits^{-\Delta}_{-\infty}d\vep'\,\mathcal{N}_S(\vep')
\left[1-f^D_\mathrm{qp}(\vep')\right]
\int\limits^{-\Delta}_{\vep'+e V_B}d\vep\,\mathcal{N}_S(\vep)
\left(\frac{1}{U_D+\vep'}-\frac{1}{U_S+\vep}\right)\approx\nonumber\\
&&\qquad\approx\frac{g_Sg_D}{2\pi^3}\,\frac{n_\mathrm{qp}}{4\nu}
\int\limits_\Delta^{\Delta-e V_B}d\vep'\,\mathcal{N}_S(\vep')
\left(\frac{1}{U_D+\vep}-\frac{1}{U_S+\Delta}+\frac{1}{U_D-\Delta}
-\frac{1}{U_S-\vep}\right).\nonumber
\end{eqnarray}
For $U_D/\Delta=-(1+\eta)$, $U_S/\Delta=\chi-(1+\eta)$, the
denominators do not vanish if $\chi-2<\eta<0$. For $\eta>0$,
the same transition can occur by a sequence of two real tunneling
processes, considered in the previous section.

For the elastic cotunneling rate we have:
\begin{eqnarray}
&&\Gamma^{\alpha\to\alpha'}_{N,\mathrm{el}}=2\sum_{k,k'}
\sum_{s_\alpha,s_{\alpha'}=\pm}
\left|\sum_n\frac{W_{\alpha'k'n}W^*_{\alpha{k}n}(1-f_n)}%
{E_N+e V_\alpha+s_\alpha\vep_k-E_{N+1}-E_n}\right|^2
\times\nonumber\\ &&\hspace*{3cm}{}\times
\frac{\vep_k+s_\alpha\xi_k}{2\vep_k}\,
\frac{1-s_\alpha(1-2f_k^\alpha)}{2}
\frac{\vep_{k'}+s_{\alpha'}\xi_{k'}}{2\vep_{k'}}\,
\frac{1+s_{\alpha'}(1-2f_{k'}^{\alpha'})}{2}
\times\nonumber\\ &&\hspace*{3cm}{}\times
2\pi\delta(e V_\alpha+s_\alpha\vep_k
-e V_{\alpha'}-s_{\alpha'}\vep_{k'}),
\end{eqnarray}
and for $N+1$ we have to replace $1-f_n\to{f}_n$ in the numerator
and put $E_{N+1}+E_n-E_N-e V_{\alpha'}-\vep_{k'}$ in the
denominator. Then, for a multi-level dot
\begin{eqnarray}
&&\Gamma^{\alpha\to\alpha'}_{\mathrm{el}}=
\frac{\delta E}{2}\,\frac{g_\alpha{g}_{\alpha'}}{(2\pi^2)^2}
\int\limits_{-\infty}^\infty{d}E\,
\mathcal{N}_S(\vep)\,d\vep\,\mathcal{N}_S(\vep')\,d\vep'\,
f^\alpha_\mathrm{qp}(\vep)\left[1-f^{\alpha'}_\mathrm{qp}(\vep')\right]
\times{}\nonumber\\&&\hspace*{3cm}{}\times
\frac{\{1-f(E),f(E)\}}{(U_\alpha+\vep-E)^2}\,
2\pi\delta(U_\alpha+\vep-U_{\alpha'}-\vep').\label{Gammael=}
\end{eqnarray}
For a single-level dot we replace
$\{1-f(E),f(E)\}\to \delta E\cdot\delta(E)$ in Eq.~(\ref{Gammael=}),
which gives
\begin{eqnarray}
&&\Gamma^{D\to S}_\mathrm{el}
\approx\frac{g_S g_D\delta E^2}{4\pi^3}\,\mathcal{N}_S(\Delta-e V_B)
\left[\frac{n_\mathrm{qp}^D/4\nu_D}{(U_D+\Delta)^2}
+\frac{n_\mathrm{qp}^S/4\nu_S}{(U_S-\Delta)^2}\right].
\end{eqnarray}

\subsection{Andreev tunneling}

There are two kinds of Andreev processes: (i) $N\to{N}+2$ with
a Cooper pair from the source electrode, and (ii)~$N+1\to{N}-1$
by injecting a Cooper pair into the drain electrode. The energies
of the initial, intermediate, and final states are
\begin{eqnarray*}
&&E_N-e V_B\to E_{N+1}+E_S+\vep_k-\frac{e V_B}2\to E_{N+2}+E_n+E_{n'}\\
&&E_{N+1}+E_n+E_n'\to E_N+E_S+\vep_k+\frac{e V_B}2\to E_{N-1}+e V_B.
\end{eqnarray*}
The matrix elements for these processes are (up to an overall sign)
\begin{eqnarray}
&&M^S_{nn'}=\sum_k\left(
\frac{W_{S k n}^*W_{S k n'}^*u_kv_k}{E_N-E_{N+1}-e V_B/2-\vep_k-E_n}
-\frac{W_{S k n}^*W_{S k n'}^*u_kv_k}{E_N-E_{N+1}-e V_B/2-\vep_k-E_{n'}}
\right)=\nonumber\\
&&\qquad{}=\phi_n^*(\vec{r}_S)\,\phi_{n'}^*(\vec{r}_S)\,
(W_S^*)^2\nu_S\left[
a\!\left(\frac{U^--E_{n'}}\Delta\right)
-a\!\left(\frac{U^--E_{n}}\Delta\right)
\right],\\
&&M^D_{nn'}=\sum_k\left(
\frac{W_{D k n}W_{D k n'}u_kv_k}{E_{N+1}-E_N-e V_B/2-\vep_k+E_n}
-\frac{W_{D k n}W_{D k n'}u_kv_k}{E_{N+1}-E_N-e V_B/2-\vep_k+E_{n'}}
\right)=\nonumber\\
&&\qquad{}=\phi_n(\vec{r}_D)\,\phi_{n'}(\vec{r}_D)\,W_D^2\nu_D
\left[a\!\left(\frac{U^++E_{n'}}\Delta\right)
-a\!\left(\frac{U^++E_{n}}\Delta\right)\right],\\
&&U^\pm\equiv{}-\frac{e V_B}2\pm(E_{N+1}-E_N),\nonumber
\end{eqnarray}
where we defined a function
\begin{eqnarray}
&&a(z)=\frac{1}{\nu_\alpha}
\sum_k\frac{|\phi_{\alpha{k}}(\vec{r}_\alpha)|^2u_kv_k}{\vep_k-z\Delta}
=\int\limits_\Delta^\infty
\frac{\vep\,d\vep}{\sqrt{\vep^2-\Delta^2}}\,
\frac{\Delta/\vep}{\vep-z\Delta}
=\int\limits_0^\infty\frac{dx}{\cosh{x}-z}
=\nonumber\\
&&\qquad{}=\frac{1}{2\sqrt{z^2-1}}
\ln\frac{-z+\sqrt{z^2-1}}{-z-\sqrt{z^2-1}}
=\frac{1}{\sqrt{z^2-1}}
\ln\frac{1-z+\sqrt{z^2-1}}{1-z-\sqrt{z^2-1}}
\mathop{\to}\limits_{z\to-\infty}\frac{\ln|2z|}{|z|}
\nonumber,\\
&&\qquad{}=\frac{\arccos(-z)}{\sqrt{1-z^2}}.
\end{eqnarray}
The rate is given by (the factor of $1/2$ from $n\leftrightarrow{n}'$
is cancelled by 2 from spin)
\begin{eqnarray}
&&\Gamma_D=(|W_D|^2\nu_D)^2\sum_{n,n'}f_nf_{n'}
|\phi_n(\vec{r}_D)|^2|\phi_{n'}(\vec{r}_D)|^2
\left[a\!\left(\frac{U^++E_{n'}}\Delta\right)
-a\!\left(\frac{U^++E_{n}}\Delta\right)\right]^2\times\nonumber\\
&&\hspace*{4cm}{}\times 2\pi\delta(E_{N+1}+E_n+E_n'-E_{N-1}-e V_B)=
\nonumber\\
&&\quad{}=\left(\frac{g_D}{4\pi^2}\right)^2\int{d}E\,dE'\,
f(E)\,f(E')\left[a\!\left(\frac{U^++E}\Delta\right)
-a\!\left(\frac{U^++E'}\Delta\right)\right]^2\times\nonumber\\
&&\hspace*{4cm}{}\times
2\pi\delta(E_{N+1}+E+E'-E_{N-1}-e V_B).
\end{eqnarray}

\subsection{Cooper-pair-electron cotunneling}

There are two types of the process:
\begin{itemize}
\item
$N+1\to{N}$, splitting a Cooper pair on $S$~electrode, leaving
a quasiparticle there, and creating a Cooper pair on $D$~electrode.
The electrostatic energy gain is $E_{N+1}+e V_S-E_N-2e V_D$.
\item
$N\to{N}+1$, splitting a Cooper pair on $S$~electrode, and
transferring a quasiparticle to $D$~electrode. The electrostatic
energy gain is $E_N+2e V_S-E_{N+1}-e V_D$.
\end{itemize}
The energies of the initial, intermediate, and final states are:
\[
\begin{array}{lcl}
E_{N+1}+E_n+2e V_S &\qquad& E_N+2e V_S\\
E_N+e V_D+\vep_{k'}+2e V_S &\quad& E_{N+1}+E_{n'}+e V_S+\vep_{k'}\\
E_{N+1}+E_{n'}+e V_D+\vep_{k'}+e V_S+\vep_k &\quad&
E_N+E_{n'}-E_{n''}+e V_D+\vep_k+e V_S+\vep_{k'}\\
E_N+2e V_D+e V_S+\vep_k+E_{n'}-E_{n''} &\quad&
E_{N+1}+E_{n'}-E_{n''}+E_n+e V_D+\vep_k
\end{array}
\]
with $n'\neq{n}''$, so they are absent in the single-level case.
In the matrix elements we take into account only intermediate states
with $N,N+1$ electrons, as states with $N-1,N+2$ electrons are
assumed to be very high in energy:
\begin{eqnarray*}
&&M_{knn'n''}^{D S D}=\sum_{k'}
\frac{W_{D k'n''}W^*_{S k n'}W_{D k'n}v_{k'}v_ku_{k'}
f_{n''}(1-f_{n'})f_n}%
{(e V_S-e V_D+E_n-E_{n'}-\vep_k-\vep_{k'})%
(E_n-\vep_{k'}-U_D)}=\nonumber\\
&&\qquad\qquad=\frac{W_D^2\nu_D\,\phi_{n}(\vec{r}_D)\,
\phi_{n''}(\vec{r}_D)W^*_{S k n'}v_k f_{n''}(1-f_{n'})f_n}%
{U_S-E_{n'}-\vep_k}\times{}\nonumber\\
&&\qquad\qquad\quad{}\times
\left[a\!\left(\frac{e V_S-e V_D+E_n-E_{n'}-\vep_k}\Delta\right)
-a\!\left(\frac{E_n-U_D}\Delta\right)\right],\\
&&M_{knn'n''}^{S D S}=\sum_{k'}
\frac{W^*_{S k'n}W_{D k n''}W^*_{S k'n'}u_{k'}u_k v_{k'}
(1-f_n)f_{n''}(1-f_{n'})}%
{(e V_S-e V_D-\vep_k-\vep_{k'}-E_{n'}+E_{n''})%
(U_S-E_{n'}-\vep_{k'})}
=\nonumber\\ &&\qquad\qquad=\frac{(W_S^*)^2\nu_S\,
\phi_{n}^*(\vec{r}_S)\,\phi_{n'}^*(\vec{r}_S)W_{S k n''}
u_k(1-f_n)f_{n''}(1-f_{n'})}{E_{n''}-\vep_k-U_D}\times{}\nonumber\\
&&\qquad\qquad\quad{}\times
\left[a\!\left(\frac{e V_S-e V_D-E_{n'}+E_{n''}-\vep_k}\Delta\right)
-a\!\left(\frac{U^S-E_{n'}}\Delta\right)\right],\\
&&U_\alpha\equiv e V_\alpha+E_N-E_{N+1}.
\end{eqnarray*}

The rates contain an additional factor of $4$ from the
spin degeneracy (note that the spins on the levels
$n,k',n''$ are locked together, and those on $n',k$ too)
\begin{eqnarray}
&&\Gamma^{D S D}_\mathrm{in}=4\,\frac{g_D^2g_S}{(2\pi)^6}
\int\limits_\Delta^\infty\mathcal{N}_S(\vep)\,d\vep
\int\limits_{-\infty}^\infty{d}E\,dE'\,dE''\,
\frac{f(E)\,[1-f(E')]f(E'')}{(U_S-E'-\vep)^2}
\times{}\nonumber\\ &&\qquad\qquad\quad{}\times
\left|a\!\left(\frac{e V_S-e V_D+E-E'-\vep}\Delta\right)
-a\!\left(\frac{E-U_D}\Delta\right)\right|^2
\times{}\nonumber\\ &&\qquad\qquad\quad{}\times
\delta\!\left(U_D+e V_D-e V_S+\vep-E+E'-E''\right),\\
&&\Gamma^{S D S}_\mathrm{in}=4\,\frac{g_Dg_S^2}{(2\pi)^6}
\int\limits_\Delta^\infty\mathcal{N}_S(\vep)\,d\vep
\int\limits_{-\infty}^\infty{d}E\,dE'\,dE''\,
\frac{[1-f(E)]\,[1-f(E')]f(E'')}{(U_D+\vep-E'')^2}
\times{}\nonumber\\ &&\qquad\qquad\quad{}\times
\left|a\!\left(\frac{e V_S-e V_D-E'+E''-\vep}\Delta\right)
-a\!\left(\frac{U^S-E'}\Delta\right)\right|^2
\times{}\nonumber\\ &&\qquad\qquad\quad{}\times
\delta\!\left(U_S+e V_S-e V_D-E-E'+E''-\vep\right).
\end{eqnarray}

For the elastic cotunneling, the processes are the same but
$n''=n'$, so in the matrix elements we should set $f_{n''}=1$
and sum over $n'$. Because the matrix element contains
$\phi_{n'}^*(\vec{r}_S)\,\phi_{n'}(\vec{r}_D)$, only one $n'$
summation survives after squaring it, so for a multilevel dot
the rates are obtained by replacing
$f(E'')\to(\delta E/2)\cdot\,\delta(E'-E'')$, where the factor
$1/2$ appears because $n'=n''$ implies also the same spin.
For a single-level dot, there is no summation over $n,n'$
either, which amounts to replacing $f(E)\,[1-f(E')]$ and
$[1-f(E)]\,[1-f(E')]$ by $\delta E^2\cdot\,\delta(E)\,\delta(E')$.

Let us estimate the zero-temperature rates in the typical regime
of turnstile operation for electron ejection (the first process),
taking $e V_B/\Delta=\chi$, $0<\chi<2$ and
$E_{N+1}-E_N=(1-\chi/2+\eta)\Delta$, with $0<\eta<\chi$.
This gives $U_D/\Delta=-(1+\eta)$, $U_S/\Delta=\chi-1-\eta$.
Then, for a multilevel dot,
\begin{eqnarray}
&&\Gamma^{D S D}_\mathrm{in}=4\Delta\,\frac{g_D^2g_S}{(2\pi)^6}
\int\limits_1^\infty\frac{w\,dw}{\sqrt{w^2-1}}
\int\limits_0^\infty\frac{dx\,dy\,dz}{(1-\chi+\eta+w+y)^2}\,
\times{}\nonumber\\ &&\qquad\quad{}\times
\left|a(\chi-w-x-y)-a(1+\eta-x)\right|^2
\delta(1+\chi+\eta-w-x-y-z),\nonumber\\
&&\Gamma^{D S D}_\mathrm{el}=2\delta E\,\frac{g_D^2g_S}{(2\pi)^6}
\int\limits_1^\infty\frac{w\,dw}{\sqrt{w^2-1}}
\int\limits_0^\infty\frac{dx\,dy}{(1-\chi+\eta+w+y)^2}\,
\times{}\nonumber\\&&\qquad\quad{}\times
\left|a(1-w-x-y)-a(1+\eta-x)\right|^2\delta(1+\chi+\eta-w-x).
\end{eqnarray}
For $\eta>0$, $1+\eta-x$ can exceed~1, then $a(1+\eta-x)$ is complex.
This corresponds to one of the intermediate states becoming real,
namely, to real ejection of the first electron into the state $k'$
on the $D$~electrode, before sending it to the condensate.

\end{document}